\documentclass[12pt,twoside]{article}
\usepackage{graphicx}
\usepackage[figuresright]{rotating}
\usepackage{myfig}

\usepackage{epsfig}
\usepackage{myfig}
\usepackage[dvips]{color}
\usepackage[english]{babel}
\usepackage{pst-node}
\usepackage{amsmath}

\definecolor{White}{rgb}{1.0,1.0,1.0}
\definecolor{Black}{rgb}{0.0,0.0,0.0}
\definecolor{Red}{rgb}{1.0,0.0,0.0}
\definecolor{Green}{rgb}{0.0,1.0,0.0}
\definecolor{Blue}{rgb}{0.0,0.0,1.0}
\definecolor{Yellow}{rgb}{1.0,1.0,0.0}
\definecolor{Magenta}{rgb}{1.0,0.0,1.0}
\definecolor{Cyan}{rgb}{0.0,1.0,1.0}
\definecolor{Orange}{rgb}{1.0,0.66,0.33}
\definecolor{Sienna}{rgb}{0.66,0.33,0.0}
\definecolor{SeaGreen}{rgb}{0.33,0.66,0.33}
\definecolor{Gray}{rgb}{0.66,0.66,0.66}
\definecolor{Maroon}{rgb}{0.33,0.33,0.33}
\definecolor{Purple}{rgb}{0.69,0.0,0.69}

\begin{document}
\title{Beam Based Alignment of Interaction Region Magnets}
\author{G.~H.~Hoffstaetter\thanks{Georg.Hoffstaetter@desy.de} and F.~Willeke\\ DESY, Hamburg, Germany}
\maketitle

\begin{abstract} 
\unitlength=1cm
\begin{picture}(0,0)
\put(12,9){\begin{minipage}{6cm}
       DESY-02-069
       \end{minipage}}
\end{picture}
In conventional beam based alignment (BBA) procedures, the relative alignment of a quadrupole to a nearby beam position monitor is determined by finding a beam position in the quadrupole at which the closed orbit does not change when the quadrupole field is varied.  The final focus magnets of the interaction regions (IR) of circular colliders often have some specialized properties that make it difficult to perform conventional beam based alignment procedures.  At the HERA interaction points, for example, these properties are:
(a) The quadrupoles are quite strong and long. 
Therefore a thin lens approximation is quite imprecise.
(b) The effects of angular magnet offsets become significant.
(c) The possibilities to steer the beam are limited as long as the alignment is not within specifications.
(d) The beam orbit has design offsets and design angles with respect to the axis of the low-beta quadrupoles.
(e) Often quadrupoles do not have a beam position monitor in their vicinity.  Here we present a beam based alignment procedure that determines the relative offset of the closed orbit from a quadrupole center without requiring large orbit changes or monitors next to the quadrupole.  Taking into account the alignment angle allows us to reduce the sensitivity to optical errors by one to two orders of magnitude.  We also show how the BBA measurements of all IR quadrupoles can be used to determine the global position of the magnets.  The sensitivity to errors of this method is evaluated and its applicability to HERA is shown.
\end{abstract}

\section{Introduction}
The new HERA interaction regions are designed to achieve a maximum possible luminosity by strongly focusing the proton beam.
This results in $\beta$-function values at the interaction point (IP) which are in the range of the bunch length. 
This new design includes superconducting combined function magnets inside the colliding beam detectors H1 and ZEUS which focus the $27.5$GeV lepton beam in the vertical plane and bend the beam away from the $920$GeV proton beam.  This allows to place the low $\beta$ magnet for the protons as close as 11m to the IP.  The synchrotron radiation produced by the beam separation has to be absorbed far away from the IP. 
Therefore the vacuum chambers downstream of the IP have a keyhole shape to allow the synchrotron radiation fan to propagate through the low--beta quadrupoles. 
These have a 28mm gap between the coils.
The aperture of the flat part of the downstream vacuum chambers is only 18mm.
This is critical because of the height of the synchrotron radiation that is generated in the upstream low--beta quadrupoles. 
Due to the large vertical divergence of the beam in these quadrupoles, the synchrotron radiation fan will only fit inside the keyhole shape if the quadrupoles in the low-beta region are aligned to a precision of better than $0.5$mm.  By optical surveying, a precision of about $0.3$mm can be achieved under optimum conditions which are not given in the interaction region (IR) with shielding walls and a large detector in between the two halves of the straight section.  Beam--based alignment was proposed to verify specifications that cannot be verified to a satisfactory precision by the survey procedure.  A precision of magnet alignment of $0.1$mm appears to be desirable. 
The magnets of the HERA IR are movable via remote control and can be adjusted in an iterative way without access to the magnets. 
\par
Beam--based alignment is a technique of deriving the position of a quadrupole magnet from the analysis of difference orbits that are generated by the variation of the strength of this quadrupole. 
If the central orbit of the beam is not in the center of the quadrupole, the beam experiences a dipole field that changes the orbit. 
Beam--position monitors detect the changes of the orbit around the ring. 
The offset of the beam with respect to the quadrupole axis is then determined by analysis of the difference orbit. 
The result may be used to calibrate the offsets of nearby beam position monitors or to mechanically re-align the quadrupole magnets.
\par

This technique has been invented to optimize the performance of the SLC
\cite{Adolphsen:1989tf,Emma:1992zz,Emma:1993en,Raimondi:1993hw}.
It has also been successfully applied to calibrate the beam position monitors in the HERA electron ring, where it was the basis for an orbit steering algorithm of minimizing the residual vertical kicks which yielded a record electron spin polarization \cite{Brinkmann:1994ff,Barber:1996hm,Boege:1996yy}.
Future accelerators such as NLC will depend heavily on extensive beam--based steering algorithms
\cite{Tenenbaum:2000ux,Tenenbaum:2001vg}.
\par
The application of beam--based alignment techniques to adjust the magnet positions in the new HERA interaction region however encountered a number of difficulties and problems. The analysis and the solutions of these problems can be helpful for future application of beam--based alignment, especially in interaction regions.
\par
The difficulties we encountered have to do with circumstances that might be considered typical for an interaction region:
(a) The quadrupoles are quite strong and long. 
Therefore the thin lens approximation is quite imprecise.
(b) The effects of angular magnet offsets become significant.
(c) The possibilities to steer the beam are limited as long as the alignment is not within specifications.
(d) The beam orbit has design offsets and design angles with respect to the axis of the low-beta quadrupoles.
(e) Often quadrupoles do not have a beam position monitor in their vicinity.
Under these circumstances the results are very sensitive to errors and it turned out to be very difficult to achieve the desired precision of the beam--based alignment of $0.1$mm. 
Moreover, since the beam cannot be centered in all the magnets simultaneously, a global analysis of the magnet positions becomes necessary which uses the results of the beam--based alignment measurements in all the IR quadrupoles.

\section{The HERA Interaction Regions}
In the following we describe the HERA Interaction region to the extent relevant for synchrotron radiation background and beam--based alignment of the low--beta quadrupoles.
\par
The proton and lepton beams collide head on in the interaction point.  The two beams are separated by combined function magnets, which start on both sides at 2m from the IP. Because of the strong synchrotron radiation power of together approximately $30$kW generated in these magnets, the layout is not symmetric. On the left side from which the lepton beam enters, there is 
a $3.2$m long relatively low field superconducting magnet
(GO) of $90$mm full aperture which deflects the leptons by $3$mrad and focuses them in the vertical plane as the first lens of a low--beta triplet.
On the right side of the IP, these functions are provided by a combination of a short (1.3m length) but large full aperture (120mm) superconducting combined function magnet (GG) and a normal conducting conventional quadrupole (GI) with a length of 1.88m. These innermost magnets are complemented by a horizontally focusing quadrupole of type GI and a vertically focusing magnet (GJ) of 1.88m length on both sides of the IP.  The double--doublet structure for focusing the protons starts at 11.2m on each side of the IP with a half quadrupole with septum plate.  Table \ref{tb:quadorb} shows the main parameters of the HERA IR quadrupoles and the location of the positron design orbit relative to the quadrupole axis in the center of each IR magnet.  For electron/proton collisions the values are slightly different.  Due to spin matching requirements it has not been possible to use exactly these design parameters for the quadrupoles, and in the routinely used optics files the computed path for injection and for luminosity operation differ by up to 0.5mm in some quadrupoles.

\begin{table}
\caption{Design parameters for the HERA IR quadrupoles and the offset of the positron design orbit in the center of these quadrupoles. \label{tb:quadorb}}
\begin{center}
\begin{tabular}{llrcrrr}
\hline
Name &
\multicolumn{1}{c}{$l$}&
\multicolumn{1}{c}{$s$}&
\multicolumn{1}{c}{focusing}&
\multicolumn{1}{c}{$k$}&
\multicolumn{1}{c}{$x_{off}$}&
\multicolumn{1}{c}{$x'_{off}$} \\
&
\multicolumn{1}{c}{[ m ]}&
\multicolumn{1}{c}{[ m ]}&
\multicolumn{1}{c}{direction}&
\multicolumn{1}{c}{[m$^{-2}$]}&
\multicolumn{1}{c}{[mm]}&
\multicolumn{1}{c}{[mrad]} \\
\hline
QL16L & 1.033 & -54.881 & y &            -0.112026 &      0 &  0     \\
QL14L & 1.033 & -42.930 & x &  \phantom{-}0.055143 &      0 &  0    \\
GJ8L  & 1.88  &  -9.172 & y &            -0.132197 & -3.288 & -0.002 \\
GI7L  & 1.88  &  -6.965 & x &  \phantom{-}0.246518 &  2.982 &  1.322 \\
GOL   & 3.20  &  -3.575 & y &            -0.140664 & -5.568 & -0.993 \\
GGR   & 1.3   &   2.625 &   &  \phantom{-}0.       &-27.913 & -1.217 \\
GI6R  & 1.88  &   4.817 & y &            -0.226504 &-10.172 &  0.165 \\
GI7R  & 1.88  &   7.218 & x &  \phantom{-}0.262005 & -0.409 & -0.358 \\
GJ8R  & 1.88  &   9.432 & y &            -0.119000 & -9.130 & -0.504 \\
QL14R & 1.033 &  43.934 & x &  \phantom{-}0.048787 &      0 &  0     \\
QL16R & 1.033 &  54.868 & y &            -0.116307 &      0 &  0     \\
\hline
\end{tabular}
\end{center}
\end{table}
\section{Analysis of Difference Orbits}
\subsection{Closed Orbit Changes due to a Quadrupole Change}
\begin{figure}[h!t!b!p!]
\begin{center}
\begin{minipage}{140mm}
\includegraphics*[width=140mm]{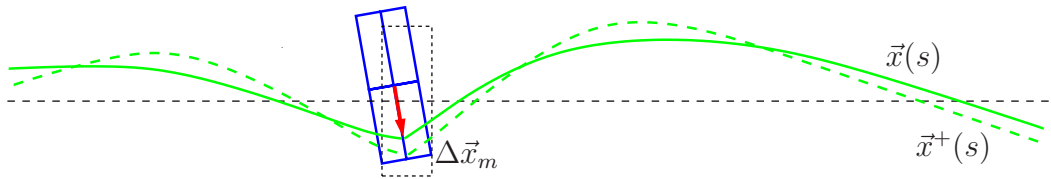}
\unitlength=1cm
\begin{picture}(0,0)
\put(5.7,0.7){$\Delta\vec x_m$}
\put(11.7,2.0){$\vec x(s)$}
\put(12.1,0.8){$\vec x^+(s)$}
\end{picture}
\end{minipage}
\caption{The orbit position relative to the axis of a quadrupole can be deduced from the closed orbit change which is created by a change in the quadrupole's field strenght.
\label{fg:bbaquad1}}%
\end{center}
\end{figure}

As described before, beam--based alignment is the analysis of difference orbits that are excited by a change in the strength of a quadrupole as illustrated in figure \ref{fg:bbaquad1}.

The difference orbit is related to the offset and the angle of the beam orbit with respect to the quadrupole axis. Therefore, we will derive the relationship between magnet alignment, closed orbit, strength variations and the difference orbit. Following standard text--book procedure, the closed orbit is written in linear approximation as
\begin{equation}
\vec{x}(s)= T_{s,0}\left(I-M_0\right)^{-1}\vec{d}_{L,0}+\vec{d}_{s,0}\ .\label{eq:xorb}
\end{equation}
In this expression, $\vec{x}(s)$ is the vector with closed orbit $x$ and it's derivative $x'$ at some longitudinal position $s$ along the design trajectory, which is chosen as reference $\vec x(s)=0$. 
$T_{s,0}$ is the transport matrix from the position $s=0$ to position 
$s$.  With the total circumference of the design curve $L$ the revolution matrix at $s$ is
$M_s=T_{s+L,s}$ and $I$ is the unity matrix.  With the focusing strength $k$, the curvature $\kappa$ of the design trajectory in the dipole fields and $\Delta\kappa$ from dipole field errors and correction coils, the vector $\vec{d}_{s,s_0}$ describes the closed orbit distortions along the ring according to the inhomogeneous equation of motion
\begin{equation}
\frac{d^2}{ds^2}d_{s,s_0} + (\kappa^2+k) d_{s,s_0}=\Delta\kappa\ \ {\rm with}\ \ d_{s_0,s_0}=0\ ,\ \ d'_{s_0,s_0}=0\ .
\end{equation} 
We assume that the quadrupole to be aligned, the test quadrupole, is the first element in the lattice.  For generality the quadrupole can also have a dipole field component.  This is important for HERA, since the magnets GG and GO are quadrupole magnets with an additional vertical dipole field.  For the beam--based alignment procedure only the quadrupole strength $k$ is changed, not the dipole field strength.  Since the quadrupole magnet has a straight axis, the motion through these fields is not correctly described with the map of a combined function magnet, where the quadrupole field is evaluated around a curved trajectory.
In the coordinate system which is aligned along the quadrupoles straight axis, the equation of particle motion through the magnetic field $B_y$ is given by
\begin{equation}
x''=-\frac{q B_y}{p}=-kx-\kappa\ .
\end{equation}
Charge and momentum are denoted by $q$ and $p$.  The transport map $\vec G(\vec x)$ which transports the phase space vector $\vec x_0$ from the beginning of the quadrupole to $\vec x_e$ at its end is given by
$\vec x_e = \vec G(\vec x_0)=G \vec x_0+\vec D$ with
\begin{equation}
G=\left(\begin{array}{cc}\cos(l\sqrt{k})&\frac{1}{\sqrt{k}}\sin(l\sqrt{k})  \\
                           -\sqrt{k}\sin(l\sqrt{k})&\cos(l\sqrt{k})\end{array}\right)\ ,\ \
\vec D=-\frac{\kappa}{k}\left(\begin{array}{c}1-\cos(l\sqrt{k})\\ \sqrt{k}\sin(l\sqrt{k})\end{array}\right)\ .
\label{eq:g0}
\end{equation}
When the alignment of the magnet with respect to the design trajectory at its entrance is described by a shift to $z_0$ and a slope $z'_0$, and similarly $\vec z_e$ describes the alignment of the end of the magnet, then the particle transport is described by
\begin{equation}
\vec x_e=G (\vec x_0-\vec z_0)+\vec z_e+\vec D\ .
\end{equation}

The closed orbit at the end of the quadrupole is given by the periodicity condition
\begin{equation}
\vec x_e=\vec G(\vec T_{L,e}(\vec x_e))=G(T_{L,e}\vec x_e+\vec d_{L,e}-\vec z_0)+\vec z_e+\vec D\ .
\label{eq:co0}
\end{equation}
After the quadrupole strength has been changed to $k+\Delta k$ we denote changed quantities by a superscript $+$, and the closed orbit is similarly given by
\begin{equation}
\vec x_e^+=\vec G^+(\vec T_{L,e}(\vec x_e^+))=G^+(T_{L,e}\vec x_e^+ +\vec d_{L,e}-\vec z_0)+\vec z_e+\vec D^+\ .
\label{eq:coplus}
\end{equation}

We are looking for a relation between the distance of the closed orbit from the quadrupole's center $\Delta \vec x=\vec x -\vec z$ and the closed orbit change $\delta \vec x=\vec x^+-\vec x$.  For this purpose we use equation (\ref{eq:co0}) to eliminate $T_{L,e}\vec x_e+\vec d_{L,e}-\vec z_0$ in equation (\ref{eq:coplus}) and obtain
\begin{eqnarray}
\vec x_e^+ &=& \delta \vec x_e+\Delta \vec x_e+\vec z_e\\
           &=& G^+(T_{L,e}\vec x_e +\vec d_{L,e}-\vec z_0+T_{L,e}\delta\vec x_e )+\vec z_e+\vec D^+\nonumber\\
           &=& G^+(G^{-1}(\Delta \vec x_e-\vec D)+T_{L,e}\delta\vec x_e )+\vec z_e+\vec D^+\ .\nonumber
\end{eqnarray}
This equation can now be solved to express $\delta \vec x$ in terms of $\Delta \vec x$,
\begin{equation}
(G^+ G^{-1}-I)\Delta \vec x_e=(I-G^+ T_{L,e})\delta\vec x_e-\vec D^+ +G^+ G^{-1}\vec D\label{eq:fulldelta}
\end{equation}
This expression can already be used to determine the magnet alignment $\Delta \vec x_e$ at its end.  However, when neglecting second orders in $\Delta k$, the expression become simplest when the alignment $\Delta\vec x_m$ in the middle of the magnet is computed.  For this we introduce the matrix $g$ and vector $\vec d$ which are $G$ and $\vec D$ in equation (\ref{eq:g0}) for half the quadrupole length.  A useful matrix will be
\begin{equation}
g^{-1}G^+g^{-1}-I=
\Delta k l\left(\begin{array}{cc}0&
\frac{1}{k}\frac{l\sqrt k-\sin(l\sqrt k)}{2l\sqrt{k}}\\
-\frac{l\sqrt k+\sin(l\sqrt k)}{2l\sqrt{k}} & 0 \end{array}\right)
+O(\Delta k^2)\ ,
\label{eq:efkicks}
\end{equation}
which reduces the effect of $\Delta k$ to the center of the quadrupole.  In the following we will use the abbreviations
\begin{equation}
\sigma^+=\frac{l\sqrt{k}+\sin{l\sqrt{k}}}{2l\sqrt{k}}\ ,\ \
\sigma^-=\frac{l\sqrt{k}-\sin{l\sqrt{k}}}{2l\sqrt{k}}\ , \ \
\underline\delta=\Delta k l
\left(\begin{array}{cc}0       &\frac{1}{k}\sigma^-\\
                       -\sigma^+&0\end{array}\right)\ .
\end{equation}
For a defocusing quadrupole ($k<0$)
$\sin$ changes to $\sinh$ due to the imaginary unit in $\sqrt{-k}$,
\begin{equation} \underline \delta
=\frac{1}{2}\frac{\Delta(-k) }{\sqrt{|k|}}
\left(
\begin{array}{cc}
           0                  &\frac{1}{|k|} [l\sqrt{|k|}-\sinh(l\sqrt{|k|})]\\
l\sqrt{|k|}+\sinh(l\sqrt{|k|})&                     0
\end{array}
\right)\ .
\end{equation}
With the revolution matrix $M_m=gT_{L,e}g$ for the middle of the test magnet equation (\ref{eq:fulldelta}) leads to
\begin{equation}
\underline\delta g^{-1}\Delta \vec x_e=(I-(I+\underline\delta) M_{m})g^{-1}\delta \vec x_e-g^{-1}(\vec D^+-\vec D)+\underline\delta g^{-1}\vec D\ .
\end{equation}
Relating $\Delta \vec x_e$ to the center of the magnet leads to $\Delta\vec x_e=g\Delta\vec x_m+\vec d$ and similarly $\vec D=g\vec d+\vec d$.  To leading order in $\Delta k$ the difference orbit around the ring $\delta \vec x(s)=T_{s,m}g^{-1}\delta\vec x_e$ is then given by

\begin{equation}
\delta\vec x(s)
=
T_{s,m}(I-M_m)^{-1}\underline\delta(\vec x_m-\vec z_m-\vec d+\Delta k\underline\delta^{-1}g^{-1}\partial_k\vec D)\ .
\end{equation}
This shows that for a quadrupole with an additional dipole field, the closed orbit distortion $\delta\vec x$ is not created by the distance $\vec x-\vec z$ between closed orbit and quadrupole axis, but by the distance between the closed orbit and an axis that is shifted by $\vec f$ from the quadrupole axis, with
\begin{equation}
\vec f = \vec d-\Delta k\underline\delta^{-1}g^{-1}\partial_k\vec D
       = \left(\begin{array}{c}f\\ 0\end{array}\right)\ {\rm with}\ 
f=\frac{\kappa}{k}
\left(\frac{1}{\sigma^+}\frac{\sin(\frac{l}{2}\sqrt{k})}{\frac{l}{2}\sqrt{k}}
-1\right)\ .\label{eq:centshift}
\end{equation}
To simplify the notation, we will now use $\Delta\vec x_m=\vec x_m-\vec z_m-\vec f$ for the closed orbit with respect to the modified axis of the quadrupole.  The shift of the alignment axis by $f$ amounts to $-401\mu m$ for the GO magnet and to $-132\mu m$ for the GG magnet.

Since $T_{s,m}(I-M_m)^{-1}$ is the closed orbit generator, the difference orbit $\delta\vec x(s)$ is created by an effective kick in the center of the test magnet.  However, there is not only an angle kick $\theta_m$ like in the thin lens model of a quadrupole, but there is also a position kick $\Delta_m$ so that the difference orbit which is created by the change of the test quadrupole has two terms,
\begin{eqnarray}
\delta x(s) &=& \delta x_\theta(s)\theta_m + \delta x_\Delta(s)\Delta_m\ ,\\
\delta x_\theta &=& T_{s,m11}(I-M_m)^{-1}_{12}+T_{s,m12}(I-M_m)^{-1}_{22}\ ,\\
\delta x_\Delta &=& T_{s,m11}(I-M_m)^{-1}_{11}+T_{s,m12}(I-M_m)^{-1}_{21}\ ,\\
\theta_m        &=& - \Delta k l\sigma^+\Delta x_m\ ,\label{eq:thetakick}\\
\Delta_m        &=&   \Delta k l\frac{\sigma^-}{k}\Delta x'_m \ .\label{eq:deltakick}
\end{eqnarray}
Contributions from angular offsets become important if 
$\frac{\sin(l\sqrt{k})}{l\sqrt{k}}$ is significantly smaller than unity, which is the case for the HERA low-beta quadrupoles as shown in table \ref{tb:thicklens}.  The well--known formulas
\begin{eqnarray}
T_{s,m11}&=&\sqrt{\frac{\beta(s)}{\beta_m}}[\cos(\phi(s)-\phi_m)+\alpha_m\sin(\phi(s)-\phi_m)]\ ,\\
T_{s,m12}&=&\sqrt{\beta(s)\beta_m}\sin(\phi(s)-\phi_m)\ ,\\
(I-M_m)^{-1}&=&\frac{1}{4\sin^2\pi \nu}
\left(\begin{array}{cc}
1-\cos2\pi \nu+\alpha_m\sin2\pi \nu & \beta_m\sin 2\pi \nu\\
-\gamma_m\sin 2\pi \nu & 1-\cos 2\pi \nu-\alpha_m\sin 2\pi \nu\end{array}\right)\nonumber
\end{eqnarray}
lead to
\begin{eqnarray}
\delta x_\theta(s) &=& \sqrt{\beta(s)\beta_m}
\frac{\cos(|\phi(s)-\phi_m|-\pi\nu)}{2 \sin(\pi\nu)}\ ,\\
\delta x_\Delta(s) &=& \sqrt{\frac{\beta(s)}{\beta_m}}
\frac{\alpha_m \cos(|\phi(s)-\phi_m|-\pi \nu)-\sin(|\phi(s)-\phi_m|-\pi \nu)}{2\sin(\pi \nu)}\ .
\end{eqnarray}
The contribution $\delta x_\theta$ is the conventional closed orbit for a correction coil at the center of the test magnet, where $\phi=\phi_m$.  The contribution $\delta x_\Delta$ can be compensated by a correction coil at $\phi_\alpha=\phi_m-{\rm atan}(\frac{1}{\alpha_m})$ since
\begin{equation}
\delta x_\Delta(s)
=
\sqrt{\beta(s)\gamma_m}\frac{\cos(|\phi(s)-\phi_\alpha|-\pi\nu)}{2\sin(\pi\nu)}{\rm sign}(\alpha_m)
\end{equation}
when $\phi(s)$ is not between $\phi_m$ and $\phi_\alpha$.  For $\phi_\alpha$ we take the branch where the atan function is in $[-\frac{\pi}{2},\frac{\pi}{2}]$.  A closed orbit correction program with correctors at these two phases will readily determin $\theta_m$ and $\Delta_m\sqrt{\frac{\gamma_m}{\beta_\alpha}}{\rm sign}(\alpha_m)$ as proposed corrector kicks.  These lead immidiately to $\Delta x_m$ and $\Delta x'_m$ with equation (\ref{eq:thetakick}) and (\ref{eq:deltakick}).

\subsection{Kick Compensation Method}
\begin{figure}[h!t!b!p!]
\begin{center}
\begin{minipage}{140mm}
\includegraphics*[width=140mm]{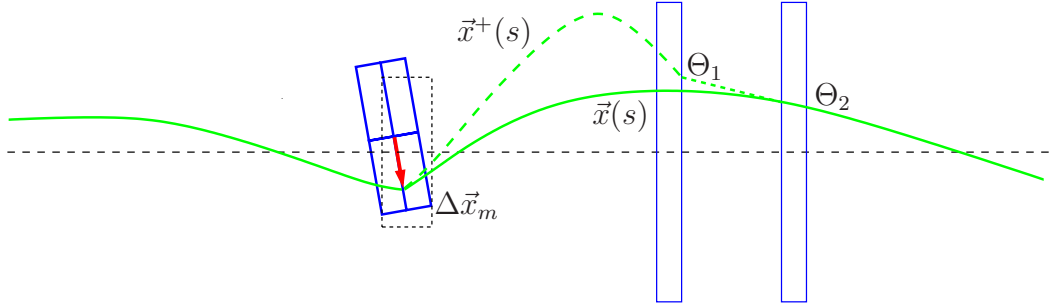}
\unitlength=1cm
\begin{picture}(0,0)
\put(5.7,1.7){$\Delta\vec x_m$}
\put(7.8,2.9){$\vec x(s)$}
\put(6.0,4.0){$\vec x^+(s)$}
\put(9.05,3.55){$\Theta_1$}
\put(10.75,3.15){$\Theta_2$}
\end{picture}
\end{minipage}
\caption{The orbit position relative to the axis of a quadrupole can be deduced from the angles required to close a bump which is exited by changing the quadrupole's field strength.
\label{fg:bbabump1}}
\end{center}
\end{figure}

Quadrupole errors around the machine might lead to a misinterpretation of the quadrupole offsets to be evaluated.  We therefore propose to create a closed bump by changing the strength of the test quadrupole by appropriately exciting two corrector coils as shown in figure \ref{fg:bbabump1}.

The difference orbit is thus a closed bump, which starts with $\delta x=0,\delta x'=0$ at the test quadrupole. The amplitude and slope $\delta\vec{x}(s)$ within this bump is derived from 
\begin{equation}
\delta\vec{x}(s)=T_{s,0} G^{-1} [(G^+-G)(\vec{x}_0-\vec{z}_0)+\vec D^+-\vec D]\ ,
\end{equation}
with the original closed orbit $\vec x_0$ and the quadrupole offset and angle alignment $\vec z_0$ at the beginning of the test quadrupole.  We again refer to the alignment in the quadrupole's center by  
\begin{eqnarray}
\delta \vec{x}(s) &=& T_{s,m} g^{-1} [ g \underline\delta g g^{-1} (\vec x_m - \vec z_m - \vec d)+\vec D^+-\vec D]\\
&=& T_{s,m}\underline \delta(\vec{x}_m-\vec z_m-\vec f)\ ,\ \ \vec f=\vec d-\Delta k\underline\delta^{-1} g^{-1}\partial_k\vec D\ .\nonumber
\end{eqnarray}
The difference orbit vanishes after the second correction coil so that
\begin{equation}
\left(\begin{array}{l}x_2\\x_2'\end{array}\right)
=
T_{s_2,m}\underline\delta\Delta\vec x_m+
T_{s_2,s_1}\left(\begin{array}{c}0\\ \theta_1\end{array}\right)+
\left(\begin{array}{c}0\\ \theta_2\end{array}\right)
\end{equation}
vanishes with $x_2=0$ and $x_2'=0$.  Here the deviation of the closed orbit from the modified quadrupole center $\Delta\vec x_m=\vec x_m-\vec z_m-\vec f$ has been used again.  The closed orbit inside the quadrupole is then
\begin{equation}
\Delta\vec x_m=-\underline\delta^{-1}[
T_{m,s_1}\left(\begin{array}{c}0\\ \theta_1\end{array}\right) +
T_{m,s_2}\left(\begin{array}{c}0\\ \theta_2\end{array}\right)]\ .
\label{eq:thetaprop}
\end{equation} 
To simplify notations, we again use $\sigma^+=[l\sqrt k+\sin(l\sqrt k)]/(2l\sqrt{k})$,  $\sigma^-=[l\sqrt k-\sin(l\sqrt k)]/(2l\sqrt{k})$ and $\phi_{1m}=\phi_1-\phi_m$. The total transformation between the compensating kicks $\vec\Theta=(\Theta_1,\Theta_2)$ and the test quadrupole offset vector is then
\begin{eqnarray}
&&\vec{x}_m-\vec{z}_m
=
\frac{\kappa}{k}
\left(\begin{array}{c}f\\ 0\end{array}\right)
+A^{-1}\vec\Theta\ ,
\label{eq:kick2x}\\
A^{-1}&=&
\frac{1}{\Delta k l}
\left(\begin{array}{cc} \sqrt{\frac{\beta_1}{\beta_m}}
\frac{\cos\phi_{1m}+\alpha_m\sin\phi_{1m}}{\sigma^+}
&
\sqrt{\frac{\beta_2}{\beta_m}}
\frac{\cos\phi_{2m}+\alpha_m\sin\phi_{2m}}{\sigma^+}
\\
\sqrt{\beta_1\beta_m}
k\frac{\sin\phi_{1m}}{\sigma^-}
&
\sqrt{\beta_2\beta_m}
k\frac{\sin\phi_{2m}}{\sigma^-}
\end{array}\right)\ ,\nonumber
\end{eqnarray}
Where the effective center shift $f$ is given in equation (\ref{eq:centshift}) and appears whenever the quadrupole field which is changed for beam based alignment is superimposed by a dipole field.

Similarly the corrector angles can be determined from the quadrupole alignment by the inverse equation,
\begin{eqnarray}
\vec\Theta
&=&
A\Delta \vec x_m\ ,\label{eq:x2kick}\\
A &=& \frac{\Delta k l}{\sin(\phi_2-\phi_1)}
\left(\begin{array}{cc} 
\phantom{-}\sqrt{\frac{\beta_m}{\beta_1}}\sin\phi_{2m}\sigma^+
&
-\frac{1}{\sqrt{\beta_1\beta_m}}
(\cos\phi_{2m}+\alpha_m\sin\phi_{2m})\frac{\sigma^-}{k}
\\
-\sqrt{\frac{\beta_m}{\beta_2}}\sin\phi_{1m}\sigma^+
&
\phantom{-}\frac{1}{\sqrt{\beta_2\beta_m}}
(\cos\phi_{1m}+\alpha_m\sin\phi_{1m})\frac{\sigma^-}{k}
\end{array}\right)\ .\nonumber
\end{eqnarray}
These formulas are accurate up to leading order in $\Delta k$.  The program MAD was used to simulate the closed orbit and the kick compensation version of beam--based alignment for the HERA IR magnets.  The inaccuracy of the reconstructed closed orbit deviation due to the neglected higher orders in $\Delta k$ was shown for all the IR magnets to be better than $1.3\%$ for $\Delta k/k\le 5\%$.  And it was shown that only second order terms in $\Delta k/k$ contribute noticeably to this small error.

\begin{figure}[h!t!b!p!]
\begin{center}
\begin{minipage}{70mm}
\includegraphics*[width=70mm]{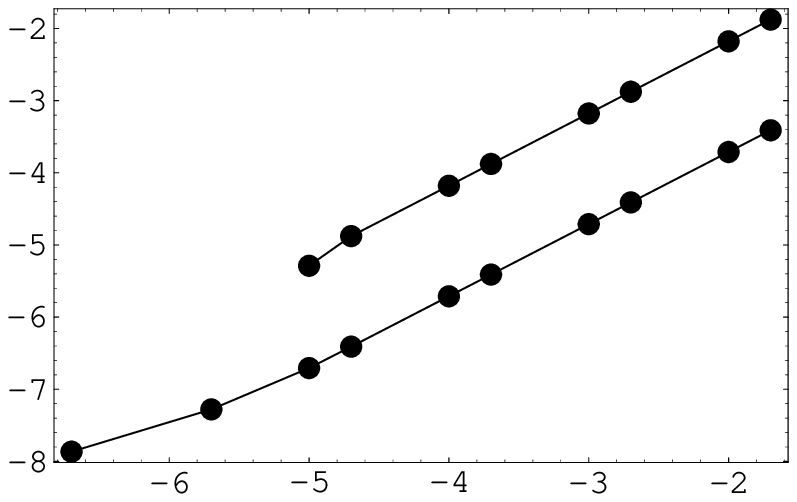}
\unitlength=1cm
\begin{picture}(0,0)
\put(0,5.1){\phantom{$\log_{10}(\frac{(x_m-z_m)_1}{x_m-z_m})$} for QL16L}
\put(5.3,0.2){$\log_{10}(\frac{\Delta k}{k})$}
\end{picture}
\end{minipage}
\begin{minipage}{70mm}
\includegraphics*[width=70mm]{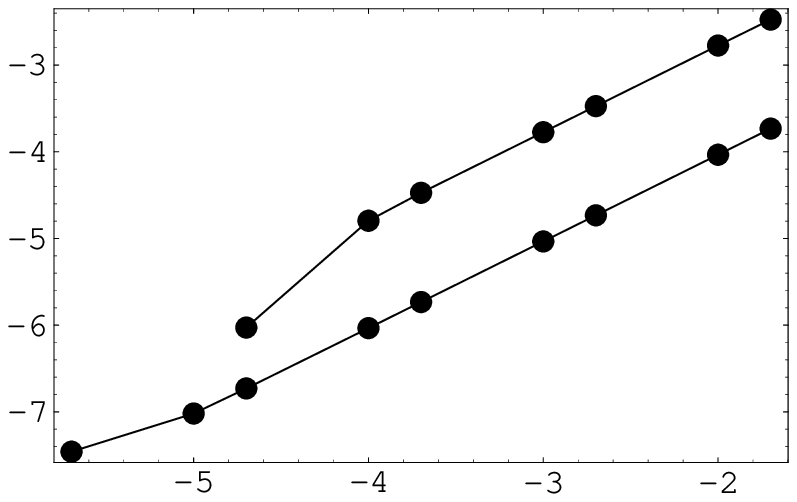}
\unitlength=1cm
\begin{picture}(0,0)
\put(0,5.1){\phantom{$\log_{10}(\frac{(x_m-z_m)_1}{x_m-z_m})$} for QL14L}
\put(5.3,0.2){$\log_{10}(\frac{\Delta k}{k})$}
\end{picture}
\end{minipage}

\vspace{8mm}

\begin{minipage}{70mm}
\includegraphics*[width=70mm]{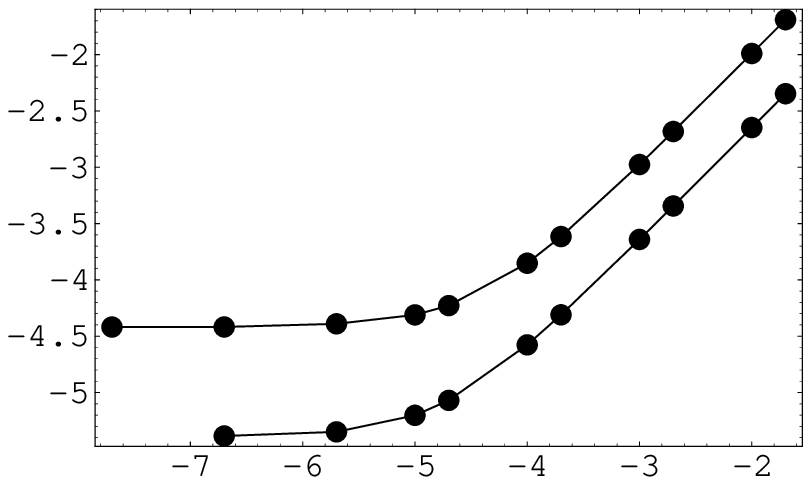}
\unitlength=1cm
\begin{picture}(0,0)
\put(0,5.1){\phantom{$\log_{10}(\frac{(x_m-z_m)_1}{x_m-z_m})$} for GOL}
\put(5.3,0.2){$\log_{10}(\frac{\Delta k}{k})$}
\end{picture}
\end{minipage}
\begin{minipage}{70mm}
\includegraphics*[width=70mm]{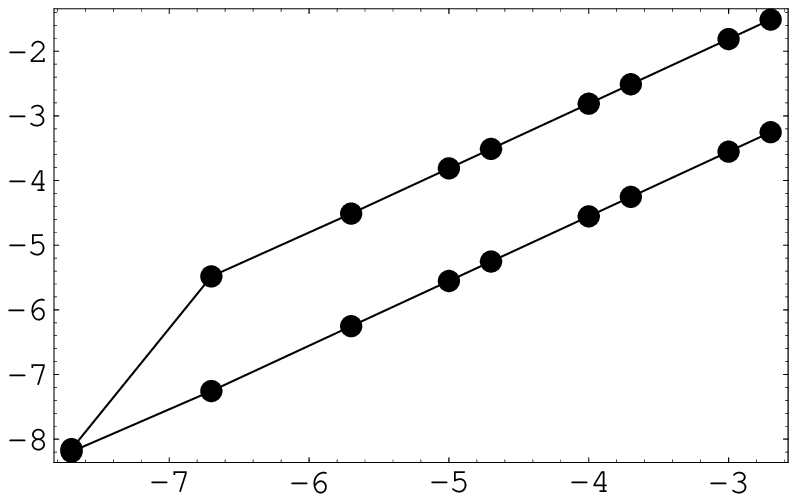}
\unitlength=1cm
\begin{picture}(0,0)
\put(0,5.1){\phantom{$\log_{10}(\frac{(x_m-z_m)_1}{x_m-z_m})$} for GGR}
\put(5.3,0.2){$\log_{10}(\frac{\Delta k}{k})$}
\end{picture}
\end{minipage}
\caption{Deviation between the alignment $\vec x_m-\vec z_m$ and the result $(\vec x_m-\vec z_m)_1$ of linearization in $\Delta k/k$ on a logarithmic scale, i.e. $\log_{10}(\frac{(x_m-z_m)_1}{x_m-z_m}-1)$ for the accuracy of position reconstuction (lower curves) and $\log_{10}(\frac{(x_m'-z_m')_1}{x_m'-z_m'}-1)$ for the accuracy of the angular reconstruction (upper curves).\label{fg:madsimerr}}
\end{center}
\end{figure}

For the quadrupoles QR16L, QR14L, GOL and GOR the error of the linearization is shown in figure \ref{fg:madsimerr}.  The deviation between the alignment $x_m-z_m$ and the first order result $(x_m-z_m)_1$ of formula (\ref{eq:kick2x}) is plotted against $\Delta k/k$ on a logarithmic scale.  The simulations were performed for the displayed range of $\Delta k/k$.  For even smaller $\Delta k/k$ numerical inaccuracies dominate the compution.  The error increases linearly, which shows that only next to leading order effects contribute noticeably to the errors and that these are small.

\begin{table}
\caption{Twiss parameters at positron injection in the center of the quadrupoles around the ZEUS IR and in the correction coils right and left of the IR which were used to create the closed bumps needed for the beam--based alignment procedure. (The strength of the IR quadrupole in the currently used injection optics deviate from the design values of table \ref{tb:quadorb} by up to 0.7\%.)\label{tb:optinj}}
\begin{center}
\begin{tabular}{lrrrrrr}
\hline
Name&
\multicolumn{1}{c}{$\beta_x$}&
\multicolumn{1}{c}{$\alpha_x$}&
\multicolumn{1}{c}{$\phi_x$}&
\multicolumn{1}{c}{$\beta_y$}&
\multicolumn{1}{c}{$\alpha_y$}&
\multicolumn{1}{c}{$\phi_y$}\\
&
\multicolumn{1}{c}{[ m ]}&
&
\multicolumn{1}{c}{[$2\pi$]}&
\multicolumn{1}{c}{[ m ]}&
&
\multicolumn{1}{c}{[$2\pi$]}\\
\hline
    CH101L & 14.609 &  0.141 & 12.550 & 14.773 & -1.571 & 11.650 \\
    CV81L  & 13.568 &  2.193 & 12.732 &  8.052 & -0.260 & 11.846 \\
    CH75L  & 14.082 & -0.608 & 12.823 &  4.426 & -0.396 & 12.055 \\
    CV56L  & 26.906 &  2.734 & 12.945 & 78.231 & -6.393 & 12.223 \\
    QL16L  & 23.463 &  1.140 & 12.950 & 86.197 & -1.784 & 12.225 \\
    QL14L  & 33.662 &  0.214 & 13.021 & 30.123 &  0.841 & 12.262 \\
    GJ8L   & 38.355 & -5.776 & 13.297 & 61.265 &  6.790 & 12.396 \\
    GI7L   & 78.538 &  1.947 & 13.303 & 23.525 &  2.619 & 12.406 \\
    GOL    & 10.899 &  3.760 & 13.322 & 27.998 &  3.462 & 12.426 \\
    GGR    &  5.040 & -1.105 & 13.614 & 19.374 & -7.296 & 12.890 \\
    GI6R   & 14.276 & -4.894 & 13.659 & 54.196 & -1.265 & 12.900 \\
    GI7R   & 62.721 & -3.665 & 13.671 & 30.056 & -0.211 & 12.910 \\
    GJ8R   & 33.574 &  4.327 & 13.679 & 60.368 & -4.849 & 12.919 \\
    QL14R  & 37.975 &  0.012 & 13.897 & 17.258 & -1.076 & 13.190 \\
    QL16R  & 23.143 & -1.007 & 13.964 & 78.119 &  0.934 & 13.242 \\
    CV56R  & 26.398 & -2.619 & 13.969 & 71.756 &  5.345 & 13.244 \\
    CH75R  & 22.662 &  0.885 & 14.066 &  8.679 &  0.509 & 13.369 \\
    CV81R  & 19.151 & -2.690 & 14.123 &  6.176 &  0.778 & 13.514 \\
    CH101R &  9.592 & -0.461 & 14.336 & 17.520 &  1.879 & 13.812 \\
\hline
\end{tabular}
\end{center}
\end{table}

The nature of this transformation becomes more transparent if the two kicks $\vec\theta=(\theta_1,\theta_2)$ are replaced by two fictive kicks $\vec\theta_f$ which occur at a betatron phase difference of exactly $\pi$ and $-{\rm atan}(1/\alpha_m)$ from the test quadrupole respectively and are normalized to the $\beta$-function.  We again take the branch where the atan function is in $[-\frac{\pi}{2},\frac{\pi}{2}]$.  With $\phi_{1m}\to\pi$ and $\phi_{2m}\to-{\rm atan}(1/\alpha_m)$ this leads to
\begin{equation}
\Delta\vec x_m
=
\frac{1}{\Delta k l}
\left(
\begin{array}{cc}
\frac{1}{\sqrt{\beta_m}}\frac{1}{\sigma^+} & 0 \\
0 & \frac{1}{\sqrt{\gamma_m}}\frac{k}{\sigma^-}
\end{array}
\right) \vec\theta_f\ .
\label{eq:thetaf2x} 
\end{equation}
The fictive angles $\vec \theta_f=(\theta_{f1},\theta_{f2})$ are now clearly related to the alignment.  The kick $\theta_{f1}$ corrects the oscillation from the magnet offset so that $\theta_{f1}=-\sqrt{\beta_m}\theta_m$ with equation (\ref{eq:thetakick}).  The kick and $\theta_{f2}$ corrects the oscillation due to the angle of the closed orbit relative to the magnet axis and $\theta_{f2}=\sqrt{\gamma_m}\Delta_m$ due to equation (\ref{eq:deltakick}).
The relation between the real and the fictive angles turns out to be
\begin{equation}
\vec{\theta}_{f}=
\left(
\begin{array}{ll}
\sqrt{\beta_1}(\cos\phi_{1,m}+\alpha_m\sin\phi_{1,m}) & 
\sqrt{\beta_2} (\cos\phi_{2,m}+\alpha_m\sin\phi_{2,m})\\
\sqrt{\beta_1} \sqrt{1+\alpha_m^2} \sin\phi_{1,m} & 
\sqrt{\beta_2} \sqrt{1+\alpha_m^2} \sin\phi_{2,m}
\end{array}
\right) \vec{\theta}\ .
\end{equation}
\section{Influence of Measurement Errors and Imperfections}
In the following, we will consider the errors in case of the compensating kick method.
\subsection{Error of Thin Lens Treatment}
Let us first consider the error that is made by treating the test quadrupole as a thin lens. The effective kick in the quadrupole of equation (\ref{eq:efkicks}) becomes in thin lens approximation 
\begin{equation}
\underline \delta^{{\rm thin}}=\left(
\begin{array}{cc}
0&0\\ -\Delta k\cdot l & 0
\end{array}
\right)\ ,
\end{equation}
leading to $-\Delta kl(x_m^{\rm thin}-z_m)=\sigma^+(x_m-z_m-f)$.
Comparing this to equation (\ref{eq:thetakick}) leads to the error of the thin lens version of beam--based alignment measurements,
\begin{equation}
x_m^{\rm thin}-z_m = \sigma^+(x_m-z_m-f)\ .
\end{equation}
The error has two components, a scaling error of $\sigma^-=1-\sigma^+$ which is shown in the 2nd column of table \ref{tb:thicklens}, and an absolute error of
$-f\sigma^+$ which is 452$\mu$m for the GOL and 132$\mu$m for the GGR magnet.

\begin{table}
\begin{center}
\caption{Measures of the accuracy of a thin kick approximation for HERA-e's interaction region quadrupoles (left) and of the relevance of angular alignment versus position alignment given by equation (\ref{eq:epsiloncomp}) .\label{tb:thicklens}}
\begin{tabular}{lccc}
\hline
Name & $\sigma^-=\frac{l\sqrt{k}-\sin(l\sqrt{k})}{2l\sqrt{k}}$ & 
\multicolumn{2}{c}{$\frac{1}{k}\frac{l\sqrt k-\sin{l\sqrt k}}{l\sqrt k+\sin{l\sqrt k}}\sqrt{\frac{\gamma_m}{\beta_m}}$}\\
     & & horiz. & vert. \\
\hline
QL16L & 0.010 & 0.006 mm/mrad & 0.002 mm/mrad \\
QL14L & 0.005 & 0.003 mm/mrad & 0.004 mm/mrad \\
GJ8L  & 0.038 & 0.044 mm/mrad & 0.032 mm/mrad \\
GI7L  & 0.069 & 0.008 mm/mrad & 0.036 mm/mrad \\
GOL   & 0.112 & 0.290 mm/mrad & 0.104 mm/mrad \\
GGR   & 0     & 0.042 mm/mrad & 0.054 mm/mrad \\
GI6R  & 0.064 & 0.100 mm/mrad & 0.009 mm/mrad \\
GI7R  & 0.073 & 0.018 mm/mrad & 0.010 mm/mrad \\
GJ8R  & 0.034 & 0.038 mm/mrad & 0.024 mm/mrad \\
QL14R & 0.004 & 0.002 mm/mrad & 0.008 mm/mrad \\
QL16R & 0.010 & 0.005 mm/mrad & 0.002 mm/mrad \\
\hline
\end{tabular}
\end{center}
\end{table}

For a HERA IR quadrupole GO the scaling error amounts to $11\%$. Given the systematic horizontal offset of $-5.5$mm in this magnet and an additional closed orbit deviation of up to 5mm, the absolute error due to thin lens analysis could be in the order of 1mm.  Also for the magnet GI in the HERA IR a thin lens evaluation could lead to an error of up to 1mm since the scaling error amounts to $6\%$ and the offset could be 10mm design offset plus a closed orbit deviation of 5mm.  The error in case of a standard lattice quadrupole with $k\approx 0.1$m$^{-2}$ and $l=1$m however is rather small. Even with an orbit offset of 5mm, the measurement error for such an element with zero design offset is only $50\mu$m.
\par  
The influence of the angle error in the quadrupole alignment
is completely ignored in the thin lens model.  To estimate the relative importance of the oscillation excited by the angle alignment, we investigate the Courant-Snyder invariant $\epsilon_{\Delta x'}$ of the part of the difference orbit which is due to the angle error and $\epsilon_{\Delta x}$ which is the part due to the quadrupole shift.  With $\Delta_m$ and $\theta_m$ from equation (\ref{eq:thetakick}) and (\ref{eq:deltakick}) we obtain
\begin{equation}
\sqrt{\frac{\epsilon_{\Delta x'}}{\epsilon_{\Delta x}}}
=
\sqrt{\frac{\gamma_m\Delta_m^2}{\beta_m\theta_m^2}}
=
\frac{1}{k}\frac{\sigma^-}{\sigma^+}
\frac{\sqrt{1+\alpha_m^2}}{\beta_m}\frac{\Delta x'}{\Delta x}
\label{eq:epsiloncomp}
\end{equation}
For the HERA IR magnets the 3rd column of table \ref{tb:thicklens} shows this ratio of oscillation amplitudes.

For the long, superconducting GO magnets in the HERA IRs the oscillation amplitude $\epsilon_{\Delta x'}$ is $29\%$ of the oscillation amplitude due to the quadrupole shift.  The quadrupole's angle and the corresponding error from neglecting it are therefore quite significant.  Thus even if the contribution of angular alignment is considered small, 
one should  take into account only that component of the difference orbit which has the proper phase relation to the test magnet in order to avoid large errors. This is especially important, if there is a large value of $\alpha_m$ in the center of the test quadrupole.
\subsection{Influence of Optical Errors}
\par
Beam optics distortions between the compensating kicks in a beam--based alignment measurement and the test quadrupole lead to misinterpretation of the difference orbit and a corresponding error of the evaluation.  Given optical errors $\delta\vec\beta=(\delta\beta_m,\delta\beta_1,\delta\beta_2)$, $\delta\alpha_m$ and $\delta\vec\phi=(\phi_{1m},\phi_{2m})$, the change $\Delta k$ in the test quadrupole requires correction kicks $\vec\theta$ to close the bumps which are obtained by inserting the perturbed optical functions in to equation (\ref{eq:x2kick}).  The inferred orbit in the quadrupole is however obtained by equation (\ref{eq:kick2x}) with the unperturbed optical functions and is therefore erroneous.  This erroneous result of the beam--based alignment procedure is here referred to as $\Delta\vec x^{\rm err}$.  We refer to the matrix in equation (\ref{eq:x2kick}) as $A(\vec\beta, \alpha_m,\vec\phi)$.  The matrix in equation (\ref{eq:kick2x}) is $A^{-1}$,
\begin{equation}
\Delta\vec x^{\rm  err}_m= A^{-1}(\vec\beta, \alpha_m,\vec\phi) A(\vec\beta+\delta\vec\beta,\alpha_m+\delta\alpha_m,\vec\phi+\delta\vec\phi)\Delta\vec x\ . 
\end{equation}

For simplicity let us now assume a phase error, so that the $\alpha$ and $\beta$ functions do not change and we assume $\delta\phi=\delta\phi_{1,m}=\delta\phi_{2,m}$ which means that no optics error occurs between the corrector magnets.
Here we will neglect all nonlinear terms in $\delta\phi$ by replacing $\cos(\delta\phi)$ by $1$ and 
$\sin(\delta\phi)$ by $\delta\phi$.  With the equations (\ref{eq:x2kick}) and (\ref{eq:kick2x}) the result can be expressed in the following way 
\begin{equation}
\Delta\vec x^{\rm err}_m-\Delta\vec x_m
=
\delta\phi\left(\begin{array}{cc}-\alpha_m & \frac{1}{k}\gamma_m\frac{\sigma^-}{\sigma^+}\\
                                 -k\beta_m\frac{\sigma^+}{\sigma^-}&\alpha_m\end{array}\right)\Delta\vec x_m\ .\label{eq:phaseerr}
\end{equation}
The term in the position error which is proportional to $\Delta x_m$ is thus simply given by
\begin{equation}
\partial_{\Delta x}\Delta x^{\rm err}_m-1=-\alpha_m\delta\phi\ .
\end{equation}
With a phase deviation $\Delta\phi=0.01\cdot 2\pi$ this error is $24\%$ for the GO quadrupole.
For the same phase deviation, table \ref{tb:phaseerr} shows all these errors for the HERA IR magnets.
\begin{table}
\begin{center}
\caption{Errors of the beam-based alignment procedure for the HERA IR quadrupoles at injection due to a phase error of $\delta\phi=0.01\cdot 2\pi$ within the closed bump of the kick compensation method.\label{tb:phaseerr}}
\begin{tabular}{lcccc}
\hline
Name &
$\partial_{\Delta x}\Delta x^{\rm err}-1$ &
$\partial_{\Delta x'}\Delta x^{\rm err}$ &
$\partial_{\Delta x}\Delta x^{' {\rm err}}$ &
$\partial_{\Delta x'}\Delta x^{' {\rm err}}-1$ \\
& & mm/mrad & mrad/mm & \\
\hline
QL16L &          -0.07 &0.00 &-16.64 &\phantom{-}0.07 \\
QL14L &          -0.01 &0.00 &-23.74 &\phantom{-}0.01 \\
GJ8L  &\phantom{-}0.36 &0.02 &-8.312 &          -0.36 \\
GI7L  &          -0.12 &0.00 &-16.29 &\phantom{-}0.12 \\
GOL   &          -0.24 &0.07 &-0.843 &\phantom{-}0.24 \\
GGR   &\phantom{-}0.07 &0.00 &-2.249 &          -0.07 \\
GI6R  &\phantom{-}0.31 &0.03 &-3.129 &          -0.31 \\
GI7R  &\phantom{-}0.23 &0.00 &-12.99 &          -0.23 \\
GJ8R  &          -0.27 &0.01 &-7.264 &\phantom{-}0.27 \\
QL14R &          -0.00 &0.00 &-26.79 &\phantom{-}0.00 \\
QL16R &\phantom{-}0.06 &0.00 &-16.42 &          -0.06 \\
\hline
\end{tabular}
\end{center}
\end{table}
The error in the position measurement that is introduced by the angle alignment is a few percent.  However, the term that generates the error in the angle determination is shown to be huge in the third column of table \ref{tb:phaseerr}.  This will prevent a precise measurement of the angular alignment.

The here studied case of a pure phase error is somewhat artificial.
When the other optical functions are also perturbed, then the evaluation becomes rather elaborate and the errors depend strongly on the location of the optical element that courses them.  We now assume that there is one thin lens quadrupole error with focal strength $\delta k_l$ at position $q$ in between the test magnet and the two correction coils.  The kicks $\vec\theta$ in the correction coils are then related to the alignment by equation (\ref{eq:thetaprop}),
\begin{equation}
\Delta\vec x_m=-\underline\delta^{-1}T_{m,q}\left(\begin{array}{cr}1 & 0\\-\delta k_l & 1\end{array}\right)[
T_{q,s_1}\left(\begin{array}{c}0\\ \theta_1\end{array}\right)+
T_{q,s_2}\left(\begin{array}{c}0\\ \theta_2\end{array}\right)]\ .
\end{equation}
The erroneously determined alignment $\Delta\vec x^{\rm err}$ does not take the optical error into account,
\begin{equation}
\Delta\vec x_m^{\rm err} =-\underline\delta^{-1}T_{m,q}[
T_{q,s_1}\left(\begin{array}{c}0\\ \theta_1\end{array}\right)+
T_{q,s_2}\left(\begin{array}{c}0\\ \theta_2\end{array}\right)]\ .
\end{equation}
We therefore obtain the relation
\begin{equation}
\Delta\vec x_m^{\rm err}-\Delta\vec x_m =\underline\delta^{-1}T_{m,q}\left(\begin{array}{cc}0 & 0\\ \delta k_l & 0\end{array}\right)
T_{q,m}\underline\delta\Delta\vec x_m\ .
\end{equation}
The error is a linear combination of the deviation $\Delta x_m$ from the magnet center and the deviation of the slope,
\begin{equation}
(\Delta x_m^{\rm err}-\Delta x_m
=
\Delta x_m(\partial_{\Delta x}\Delta x^{\rm err}_m-1)
+
\Delta x'_m\partial_{\Delta x'}\Delta x^{\rm err}_m\ .
\end{equation}
and similarly for the error of the angular alignment determination.
The exact value of the terms in the matrix that relates $\Delta\vec x^{\rm err}$ and $\Delta\vec x$ depend on the optical parameters, especially on the phase advance between the error and the test quadrupole.  When one inserts as a worst case scenario for each of the matrix elements the phase $\phi_{qm}$ where it has the maximum absolute value, one obtains
\begin{eqnarray}
{\rm Max}|\partial_{\Delta x}\Delta x^{\rm err}_m-1|
&=&
|\delta k_l| \beta_q\frac{1}{2}(|\alpha_m|+\sqrt{1+\alpha_m^2})\ ,
\nonumber\\
{\rm Max}|\partial_{\Delta x'}\Delta x^{\rm err}_m|
&=&
|\delta k_l| \beta_q\gamma_m\frac{\sigma^-}{k\sigma^+}\ ,
\nonumber\\
{\rm Max}|\partial_{\Delta x}\Delta x^{' {\rm err}}_m|
&=&
|\delta k_l| \beta_q\beta_m\frac{k\sigma^+}{\sigma^-}\ ,
\nonumber\\
{\rm Max}|\partial_{\Delta x'}\Delta x^{' {\rm err}}_m-1|
&=&
|\delta k_l| \beta_q\frac{1}{2}(|\alpha_m|+\sqrt{1+\alpha_m^2})\ .
\label{eq:maxquaderr}
\end{eqnarray}
These values are shown for the HERA IR in table \ref{tb:phaseerr} when a thin lens quadrupole error with tune change $\frac{1}{4\pi}\delta k_l\beta_q = 0.01$ is assumed.  Table \ref{tb:tuneerr} shows which relative quadrupole errors in the IR lead to such a tune shift.
\begin{table}
\begin{center}
\caption{Maximum of the errors of the beam-based alignment procedure for the HERA IR quadrupoles at injection due to some focusing error at position $s_q$ with tune shift $\frac{1}{4\pi}\beta_q\delta k_l=0.01$.  Corresponding quadrupole errors are shown in table \ref{tb:tuneerr}.\label{tb:maxquaderr}}
\begin{tabular}{lcccc}
\hline
Name &
$|\partial_{\Delta x}\Delta  x^{\rm err}-1|$&
$|\partial_{\Delta x'}\Delta  x^{\rm err}|$&
$|\partial_{\Delta x}\Delta x^{' {\rm err}}|$&
$|\partial_{\Delta x'}\Delta x^{' {\rm err}}-1|$\\
& & mm/mrad & mrad/mm & \\
\hline
QL16L & 0.17 & 0.00 & 33 & 0.17 \\
QL14L & 0.08 & 0.00 & 47 & 0.08 \\
GJ8L  & 0.73 & 0.03 & 17 & 0.73 \\
GI7L  & 0.26 & 0.00 & 33 & 0.26 \\
GOL   & 0.48 & 0.14 & 2. & 0.48 \\
GGR   & 0.16 & 0.01 & 4. & 0.16 \\
GI6R  & 0.62 & 0.06 & 6. & 0.62 \\
GI7R  & 0.47 & 0.01 & 25 & 0.47 \\
GJ8R  & 0.55 & 0.02 & 14 & 0.55 \\
QL14R & 0.06 & 0.00 & 53 & 0.06 \\
QL16R & 0.15 & 0.00 & 32 & 0.15 \\
\hline
\end{tabular}
\end{center}
\end{table}
\begin{table}
\begin{center}
\caption{Relative filed strength errors for the IR quadrupoles which lead to a tune shift $\Delta Q_x$ or $\Delta Q_y$ of $0.01$.\label{tb:tuneerr}}
\begin{tabular}{lccccccccccc}
\hline
Name &
QL16L & QL14L & GJ8L & GI7L & GOL \\
\hline
$\frac{4\pi\Delta Q_x}{\beta_x kl}$ &
4.6\% & 6.5\% & 1.3\% & 0.3\% & 2.6\% \\
$\frac{4\pi\Delta Q_y}{\beta_y kl}$ &
1.3\% & 7.3\% & 0.8\% & 1.1\% & 1.0\% \\
\hline\hline
Name &
GI6R & GI7R & GJ8R & QL14R & QL16R \\
\hline
$\frac{4\pi\Delta Q_x}{\beta_x kl}$ &
2.1\% & 0.4\% & 1.7\% & 6.6\% & 4.5\%\\
$\frac{4\pi\Delta Q_y}{\beta_y kl}$ &
0.5\% & 0.8\% & 0.9\% & 14\% & 1.3\%\\
\hline
\end{tabular}
\end{center}
\end{table}

For specific quadrupole errors in the HERA interaction region, the sensitivity of this beam based alignment procedure was also evaluated.  Table \ref{tb:quaderr} shows the error of the constructed beam offset in the GJ8L, GI7L, and GOL magnet which occurs when the field strength in one of the other quadrupoles on the left side of the IP has an error which leads to a tune shift of 0.01.  We only show the most relevant term of the error, $\partial_{\Delta x}(\Delta  x^{\rm err}-\Delta x )$.

\begin{table}
\begin{center}
\caption{Horizontal errors $\partial_{\Delta x}x^{\rm err}-1$ of the beam-based alignment procedure for three HERA IR quadrupoles at injection due to a error of the field strength in one of the other IR quadrupoles which leads to a 0.01 tune shift.  All magnets were assumed to be correctly aligned.\label{tb:quaderr}}
\begin{tabular}{l|ccccc}
\hline
error element &
for GJ8L & for GI7L & for GOL \\
\hline
QL16L & -0.432 &  0.217  &  0.309 \\
QL14L & -0.687 &  0.259  &  0.463 \\
GJ8L  &        & -0.005  & -0.008 \\
GI7L  &        &         & -0.008 \\
\hline
\end{tabular}
\end{center}
\end{table}

\subsection{Reduction of Sensitivity to  Errors}
One source of errors is an imperfect determination of the compensation kicks $\theta_1$ and $\theta_2$.  Since the determination of $\Delta x'_m$ is very prone to errors, as can be seen in the tables \ref{tb:phaseerr} and \ref{tb:maxquaderr}, it is not worth trying to determine the angle alignment.  But we will assume that the angle alignment $\Delta x'_m=x'_m-z'_m$ is approximately correct and we will therefore require our compensation to lead to the design value $z_m^{' 0}$ of the angular alignment.  While the angles $\vec\Theta$ are measured, we assume that the correct angles to close the bump would have been $\vec\Theta-\Delta\vec\Theta$.  Since the errors $\Delta\vec\Theta$ are not known, we introduce an estimate $\Delta\vec\Theta^*$ of the erroneous angle such that equation (\ref{eq:kick2x}) leads to an estimated alignment of 
\begin{equation}
\Delta \vec x_m^*
=
A^{-1}(\vec\beta,\alpha_m,\vec\phi)[\vec\Theta-\Delta\vec\Theta^*]
=
\left(\begin{array}{c}\Delta x_m^* \\ -z_m^{' 0}\end{array}\right)\ .
\end{equation}
With $\vec a_2=(A^{-1}_{2,1},A^{-1}_{2,2})$ we write $\vec a_2\cdot[\vec\Theta-\Delta\vec\Theta^*]=-z_m^{' 0}$.  This condition should be satisfied for a set $\Delta\vec\Theta^*$ of angles which is as small as possible, i.e. $|\Delta\vec\Theta^*|^2$ should be minimal.  We can use Lagrange multipliers to minimize,
\begin{eqnarray}
\Delta\vec\Theta^{*2}+\lambda[\vec a_2\cdot(\vec\Theta-\Delta\vec\Theta^*)+z_m^{' 0}]
&\to&{\rm \ minimum}\ ,\\
2\Delta{\vec\Theta}^*-\lambda\vec a_2&=&0\ ,\\
\vec a_2\cdot\Delta\vec\Theta^*
&=&\vec a_2\cdot\vec\Theta+z_m^{' 0}\ .
\end{eqnarray}
These equations lead to $\Delta\vec\Theta^*=\vec a_2(\vec a_2\cdot\vec\Theta+z_m^{' 0})/|\vec a_2|^2$.  With $\vec a_1=(A^{-1}_{11},A^{-1}_{12})$ equation (\ref{eq:kick2x}) determins the alignment to $\Delta x_m=\vec a_1\cdot\vec\Theta$.  When the above estimate $\Delta\vec\Theta^*$ is used, the estimated alignment is given by
\begin{eqnarray}
\Delta x_m^*
&=&\frac{1}{|\vec a_2|^2}[|\vec a_2|^2\vec a_1^T-(\vec a_1\cdot\vec a_2)\vec a_2^T]\vec\Theta-\frac{\vec a_1\cdot\vec a_2}{|\vec a_2|^2}z_m^{' 0}\nonumber\\
&=& \frac{1}{|\vec a_2|^2}\vec a_2^T(\vec a_2,-\vec a_1)
\left(\begin{array}{c}\vec a_1^T\\ \vec a_2^T\end{array}\right)\vec\Theta
-\frac{\vec a_1\cdot\vec a_2}{|\vec a_2|^2}z_m^{' 0}\nonumber\\
&=&
\frac{1}{|\vec a_2|^2}\vec a_2^T
\left(\begin{array}{cc}A^{-1}_{21}&-A^{-1}_{11}\\A^{-1}_{22}&-A^{-1}_{12}\end{array}\right)
\left(\begin{array}{cc}A^{-1}_{11}&A^{-1}_{12}\\A^{-1}_{21}&A^{-1}_{22}\end{array}\right)\vec\Theta
-\frac{\vec a_1\cdot\vec a_2}{|\vec a_2|^2}z_m^{' 0}\nonumber\\
&=&
\frac{{\rm det}(A^{-1})}{|\vec a_2|^2}\vec a_2^T
\left(\begin{array}{cr}0&-1\\1&0\end{array}\right)\vec\Theta
-\frac{\vec a_1\cdot\vec a_2}{|\vec a_2|^2}z_m^{' 0}\ .
\label{eq:kick2xcorr}\\
\end{eqnarray}
With equation (\ref{eq:kick2x}) for $A^{-1}$ this yields
\begin{eqnarray}
\Delta x_m^*
&=&
\frac{\sqrt{\beta_1\beta_2}\sin\phi_{21}}{\Delta k l\sigma^+\sqrt{\beta_m}}
\frac{\sqrt{\beta_2}\sin\phi_{2m}\Theta_1-\sqrt{\beta_1}\sin\phi_{1m}\Theta_2}
{\beta_1\sin^2\phi_{1m}+\beta_2\sin^2\phi_{2m}}\\
&-&
z_m^{' 0}\frac{\sigma^-}{k\beta_m\sigma^+}(\alpha_m+
\frac{\beta_1\sin\phi_{1m}\cos\phi_{1m}+\beta_2\sin\phi_{2m}\cos\phi_{2m}}
{\beta_1\sin^2\phi_{1m}+\beta_2\sin^2\phi_{2m}})
\ .\nonumber
\end{eqnarray}

\begin{figure}[h!t!b!p!]
\begin{center}
\begin{minipage}{140mm}
\includegraphics*[width=140mm]{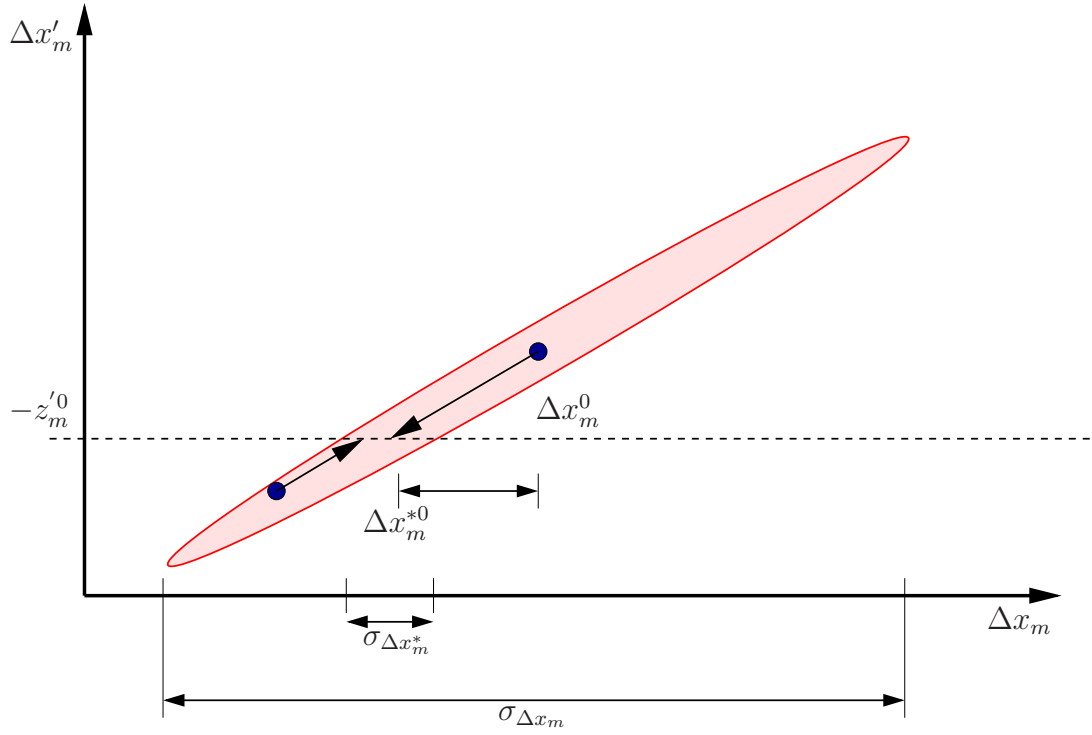}
\unitlength=1cm
\begin{picture}(0,0)
\put(6.0,0.5){$\sigma_{\Delta x_m}$}
\put(4.2,1.5){$\sigma_{\Delta x_m^*}$}
\put(6.5,4.5){$\Delta x_m^0$}
\put(4.2,3.0){$\Delta x_m^{*0}$}
\put(12.5,1.7){$\Delta x_m$}
\put(-0.5,9.5){$\Delta x_m'$}
\put(-0.5,4.5){$-z_m^{' 0}$}
\end{picture}
\end{minipage}
\caption{The reduction of the spread of the determined offset $\Delta x_m^*$ and the introduced systematic error $\Delta x_m^{*0}-\Delta x_m^0$ due to the requirement of $\Delta x_m^{'0}=-z_m^{' 0}$.\label{fg:errorreduce}}
\end{center}
\end{figure}

When the determination of the angles $\vec\Theta$ has an error with standard deviation $\sigma_\Theta$, then the errors in the determination of $\Delta x_m$ and $\Delta x_m'$ have the standard deviations $|\vec a_1|\sigma_\Theta$ and $|\vec a_2|\sigma_\Theta$ when equation (\ref{eq:kick2x}) is used, leading to
\begin{equation}
\sigma_{\Delta x_m}=\frac{\sigma_\Theta}{\Delta k l\sigma^+}\sqrt{\frac{\beta_1}{\beta_m}(\cos\phi_{1m}+\alpha_m\sin\phi_{1m})^2+\frac{\beta_2}{\beta_m}(\cos\phi_{2m}+\alpha_m\sin\phi_{2m})^2}\ .\label{eq:sigmax}
\end{equation}
When the equation (\ref{eq:kick2xcorr}) is used, the standard deviation of $\Delta x_m^*$ is always smaller,
\begin{eqnarray}
\sigma_{\Delta x_m^*}
&=&\sigma_\Theta\left|\frac{{\rm det}(A^{-1})}{|\vec a_2|^2}\vec a_2^T
\left(\begin{array}{cr}0&-1\\1&0\end{array}\right)\right|\label{eq:errorcor}\\
&=&
\frac{\sigma_\Theta}{\Delta k l\sigma^+\sqrt{\beta_m}}\frac{\sqrt{\beta_1\beta_2}\sin\phi_{21}}
{\sqrt{\beta_1\sin^2\phi_{1m}+\beta_2\sin^2\phi_{2m}}}\ .\nonumber
\end{eqnarray}
Especially for large $\alpha_m$ his spread of results in $\Delta x_m^*$ is drastically smaller than the spread in equation (\ref{eq:sigmax}).
If the angular alignment of the orbit relative to the magnet is not the design value $-z_m^{'0}$, then $\Delta x_m^*$ contains a systematic error since $\Delta\Theta=0$ does not lead to the correct alignment $\Delta x_m=\vec a_1\cdot\vec\Theta$ but with $\vec a_2\cdot\vec\Theta=\Delta x_m'$ it leads to
\begin{eqnarray}
\Delta x_m^*
&=&
\Delta x_m-(\Delta x_m'+z_m^{' 0})
\frac{\vec a_1\cdot\vec a_2}{|\vec a_2|^2}\\
&=&
\Delta x_m-(x_m'-z_m'+z_m^{' 0})\times\nonumber\\
&&
\frac{\sigma^-}{k\sigma^+\beta_m}(\alpha_m+
\frac{\beta_1\cos\phi_{1m}\sin\phi_{1m}-\beta_2\cos\phi_{2m}\sin\phi_{2m}}{\beta_1\sin^2\phi_{1m}+\beta_2\sin^2\phi_{2m}})\ .
\end{eqnarray}
This systematic deviation of $\Delta x_m^{*0}=\Delta x_m^*(\Delta\vec\Theta=0)$ from $\Delta x_m^{0}=\Delta x_m(\Delta\vec\Theta=0)$ is shown in tabel \ref{tb:systerror} for an angular deviation $x'_m$ of 1mrad.  Figure \ref{fg:errorreduce} shows that a spherical error distribution for $\Theta_1$ and $\Theta_2$ leads to an elliptical distribution for $\Delta x_m$ and $\Delta x_m'$.  The large spread in $\Delta x_m$ is reduced by the estimation of $\Delta\vec\Theta^*$ as shown in the figure.  This, however, introduces the systematic error $x_m^{*0}-x_m^0$ which is also shown.

\begin{table}
\begin{center}
\caption{Systematic error $\Delta x_m^{*0}-\Delta x_m^0$ of the procedure which reduces the measurements sensitivity to the kick angles $\vec\Theta$.
This error increases linearly with the deviation from the orbits design angle in the magnet.
\label{tb:systerror}}
\begin{tabular}{ccccccc}
\hline
GJ8L     & GI7L  & GOL     & GGR     & GI6R    & GI7R    & GJ8R    \\ \hline
-43 $\frac{\mu m}{mrad}$ &
  8 $\frac{\mu m}{mrad}$ &
291 $\frac{\mu m}{mrad}$ &
-34 $\frac{\mu m}{mrad}$ &
-99 $\frac{\mu m}{mrad}$ &
-18 $\frac{\mu m}{mrad}$ &
 37 $\frac{\mu m}{mrad}$ \\
\hline
\end{tabular}
\end{center}
\end{table}

Now we will show that this procedure also reduces the sensitivity to optical errors.   In the tables \ref{tb:phaseerr}, \ref{tb:maxquaderr} and \ref{tb:quaderr} is has been seen that the most important error of the alignment determination is due to the term $\partial_{\Delta x}\Delta x^{\rm err}-1$. This is mostly due to the fact that the Twiss parameter $\alpha_m$ that contributes to this term can be relatively large.  We will now show that the here proposed method of error reduction makes this term independent of $\alpha$ for all types of optical errors.  For an alignment $\Delta x_m$, the angles $\vec\Theta$ that close the bump are given by $\vec\Theta=A(\vec{\tilde\beta},\tilde\alpha_m,\vec{\tilde\phi})\Delta\vec x_m$
where the tilde indicates Twiss parameters which are perturbed due to an optical error.  With equation (\ref{eq:errorcor}) the estimate of the alignment is computed by
\begin{equation}
\Delta x_m^*
=
\frac{{\rm det}(A^{-1})}{|\vec a_2|^2}(A^{-1}_{22},-A^{-1}_{21})\vec\Theta
-
\frac{\vec a_1\cdot\vec a_2}{|\vec a_2|^2}z_m^{' 0}\ .
\end{equation}
The most disturbing error contribution $\partial_{\Delta x}\Delta x_m^{\rm err}-1$ is then given by
\begin{equation}
\partial_{\Delta x}\Delta x_m^{\rm err}-1
=
\frac{{\rm det}(A^{-1})}{|\vec a_2|^2}
\left(\begin{array}{r}
\frac{\sqrt{\beta_m\beta_2}}{\sigma^-}\sin\phi_{2m}\\
-\frac{\sqrt{\beta_m\beta_1}}{\sigma^-}\sin\phi_{1m}
\end{array}\right)
\cdot
\left(\begin{array}{r}
\sqrt{\frac{\tilde\beta_m}{\tilde\beta_1}}
\frac{\sin\tilde\phi_{2m}}{\sin(\tilde\phi_{21})}\sigma^+\\
-\sqrt{\frac{\tilde\beta_m}{\tilde\beta_2}}
\frac{\sin\tilde\phi_{1m}}{\sin(\tilde\phi_{21})}\sigma^+
\end{array}\right) \ .\label{eq:toterrcor}
\end{equation}
Since det($A^{-1}$) as well as $|\vec a_2|^2$ do not depend on $\alpha_m$,
the error contribution no longer depends on $\alpha_m$, no matter which optical perturbation occurs.  Equation (\ref{eq:phaseerr}) and table \ref{tb:phaseerr} show the error terms introduced by an optical error that only changes the betatron phase.  When the error reduction method is used, the error of the alignment determination can be computed from equation (\ref{eq:toterrcor}) to be
\begin{equation}
\partial_{\Delta x}\Delta x_m^{\rm err}-1
=
2\frac{\beta_1\sin 2\phi_{1m}+\beta_2\sin 2\phi_{2m}}{\beta_1\sin^2\phi_{1m}+\beta_2\sin^2\phi_{2m}}\delta\phi\ .
\end{equation}
Also the term $\partial_{\Delta x}\Delta x_m^{\rm err}-1$ depends on the Twiss parameters at the corrector coils.  For $\phi_{1m}=\pi/2$ and $\phi_{2m}=\pi$ we obtain
\begin{equation}
\Delta x^*=\delta\phi\frac{\sigma^-}{k\beta_m\sigma^+}\Delta x'_m\ .
\end{equation}
The error becomes completely independent of $\Delta x_m$ and the already small error due to $\Delta x_m'$ in equation (\ref{eq:phaseerr}) is reduced by $\sqrt{1+\alpha_m^2}$.  For the phase advances which are realized in the 
HERA IR between the test magnet and two horizontal correction coils at 101m and 75m left of the IP, the error terms are shown in table \ref{tb:phaseerrcor}.  The error has been reduced to less than 1\% for all magnets, whereas it was up to 36\% without error reduction.

\begin{table}
\begin{center}
\caption{After error reduction: the horizontal beam-based alignment procedure for the HERA IR quadrupoles at injection due to a phase error of $\delta\phi=0.01\cdot 2\pi$ within the closed bump of the kick compensation method.\label{tb:phaseerrcor}}
\begin{tabular}{lcccc}
\hline
Name &
$\partial_{\Delta x}\Delta x^{\rm err}-1$ &
$\partial_{\Delta x'}\Delta x^{\rm err}$ &
Max$|\partial_{\Delta x}\Delta x^{\rm err}-1|$ &
Max$|\partial_{\Delta x'}\Delta x^{\rm err}|$ \\
& & mm/mrad & & mm/mrad \\
\hline
QL16L &          -0.002 & 0.000 & 0.063 & 0.000 \\
QL14L &          -0.005 & 0.000 & 0.063 & 0.000 \\
GJ8L  &\phantom{-}0.009 & 0.001 & 0.063 & 0.003 \\
GI7L  &\phantom{-}0.009 & 0.000 & 0.063 & 0.001 \\
GOL   &\phantom{-}0.009 & 0.002 & 0.063 & 0.023 \\
GGR   &          -0.007 & 0.002 & 0.063 & 0.004 \\
GI6R  &          -0.003 & 0.001 & 0.063 & 0.008 \\
GI7R  &          -0.002 & 0.003 & 0.063 & 0.001 \\
GJ8R  &          -0.001 & 0.001 & 0.063 & 0.003 \\
QL14R &\phantom{-}0.005 & 0.000 & 0.063 & 0.000 \\
QL16R &          -0.004 & 0.000 & 0.063 & 0.000 \\
\hline
\end{tabular}
\end{center}
\end{table}

The maximum error terms that can occur due to a focusing error somewhere in the bump depend also on the phase advances when the error reduction method is used.  For $\phi_{1m}=\pi/2$ and $\phi_{2m}=\pi$ one obtains
\begin{eqnarray}
{\rm Max}|\partial_{\Delta x}\Delta x^*_m-1|
&=&
\frac{\delta k_l\beta_q}{2}\ ,\\
{\rm Max}|\partial_{\Delta x'}\Delta x^*_m|
&=&
\frac{\delta k_l\beta_q(1+\sqrt{1+\alpha_m^2})}{2}\frac{\sigma^-}{k\beta_m\sigma^+}\ .
\end{eqnarray}
Both terms are always smaller than the maximum errors in equation (\ref{eq:maxquaderr}) without error reduction.  Especially the first error term is significantly smaller as can be seen in table \ref{tb:phaseerrcor} where these maximum errors are plotted for the IR of HERA.  In table \ref{tb:quaderrcor} the error for one of the IR magnets is shown which occurs when another IR magnet causes the focusing error.  With error reduction also these errors are significantly smaller than those in table \ref{tb:quaderr}.

\begin{table}
\begin{center}
\caption{After error reduction: $\partial_{\Delta x}\Delta x^{\rm err}-1$ for three HERA IR quadrupoles at injection due to a error of the field strength in one of the other IR quadrupoles which leads to a 0.01 tune shift.\label{tb:quaderrcor}}
\begin{tabular}{lccccc}
\hline
error element &
for GJ8L &
for GI7L &
for GOL \\
\hline
QL16L &  0.048 &  0.049  &  0.053 \\
QL14L &  0.004 &  0.008  &  0.021 \\
GJ8L  &        & -0.005  & -0.020 \\
GI7L  &        &         & -0.015 \\
\hline
\end{tabular}
\end{center}
\end{table}

We conclude the error considerations by realizing that there are no chances
to determine the angular alignment of the quadrupoles with the desired precision of $\Delta x_m'\simeq 100\mu$rad.
The large sensitivity to optical errors and to corrector settings of the estimated quadrupole offset can be reduced drastically by one to two orders of magnitude.
\section{The Global Positions of Magnets}
\subsection{Combining BBA Data of all IR Magnets}
Since it is not possible to steer the beams to the middle of all  quadrupoles for a misaligned interaction region, the measurement of the position and angle of the beam with respect to a single quadrupole magnet does not give enough information to determine the global alignment of this magnet.
All the quadrupole offsets and angles with respect to the beam have to be determined, and the beam orbit has to be consistently modeled, thereby fixing the absolute magnet positions. 
In order to achieve that, the following procedure has been established: the beam offsets with respect to all the quadrupoles in the IR are measured for two or more different quadrupole settings in the IR.  Then a model of the IR that has the initial orbit values at the entrance of the IR and the magnet positions as free parameters is fitted to the set of measurements. An additional constraint in the fit is that the magnet position deviations from their nominal values should be minimal in order to connect to the machine coordinate system avoiding a global, unrealistic offset. 
\par
In order to perform this task, we need an explicit formula for the beam orbit as a function of alignments and initial conditions.
To arrive at such a formula we write the transformation of the beam orbit from the 
center of a quadrupole to the center of its neighbor quadrupole as
\begin{equation}
\vec{x}_n= g_n \left\{D_{n,n-1}
          [g_{n-1}(\vec{x}_{n-1}-\vec{z}_{n-1}) + \vec d_{n-1} + t_{n-1}\vec{z}_{n-1}]+\vec d_{n,n-1}
          -t_n^{-1}\vec{z}_n\right\} + \vec d_n + \vec{z}_n\ .
\nonumber
\end{equation}
The $4\times 4$ matrix $D_{n,n-1}$ is the transport matrix from the end of the n-1st test magnet to the entrance of the $n$th test magnet.  On this distance the closed orbit distortions $\vec d_{n,n-1}$ due to corrector coils or field errors are being accumulated.  The matrix $t_n$ describes a drift with half the length of the $n$th magnet.  It is used to obtain the alignment at the end of a magnet as $t_{n-1}\vec z_{n-1}$ or at the beginning of a magnet as $t^{-1}_n\vec z_n$.  In the following it will use the $4\times 4$ matrix $T_{n,n-1} = g_n D_{n,n-1} g_{n-1}$.  The matrix $T_{n,n-1}$ transforms from center to center between two neighbored quadrupoles.  The vector $\vec{x}_n$ describes the orbit and $\vec{z}_n$ is the vector of magnet alignments, both taken in the center of the magnet, $g_n$ transforms through half the quadrupole with index n.  This equation can be simplified by using $5\times 5$ matrixes, where the fifth column is used to describe the closed orbit distortions.  The orbit vector then has five components, $(x,x',y,y',1)$, the alignment vectors $\vec z_n$ have $0$ in their 5th component.  The closed orbit deviations $\vec d$ are then all absorbed in the 5th columns so that, after combining the terms, the recursive orbit formula reads 
\begin{equation}
\vec{x}_n= T_{n,n-1}
\vec{x}_{n-1}-T_{n,n-1}(I-g_{n-1}^{-1} t_{n-1})\vec{z}_{n-1}         
 +(1-g_n t_n^{-1})\vec{z}_n\ .
\end{equation}
This recursive formula leads to the explicit expression
\begin{eqnarray}
\vec{x}_n &=& T_{n,0}\vec{x}_0 - T_{n,0}(I-g_0^{-1}t_0)\vec{z}_0 
\label{eq:xproper}\\
          &+&\sum_{j=1}^{n-1} T_{n,j} ( g_j^{-1} t_j - g_j t_j^{-1}) \vec{z}_j
+(I-g_n t_n^{-1})\vec{z}_n\ .\nonumber
\end{eqnarray}
Using 
\begin{eqnarray}
P_{n,0} &=& -T_{n,0}(I-g_0^{-1} t_0)\ ,\\
P_{n,j} &=&  T_{n,j} (g_j^{-1} t_j- g_j t_j^{-1})\ {\rm for}\ 0<j<n\ ,\\
P_{n,n} &=& -g_n t_n^{-1}\ .
\end{eqnarray}
We finally obtain for the orbit in each of the $N$ test magnets the desired form
\begin{equation}
\vec{x}_n-\vec{z}_n= T_{n,0}\vec{x}_0+\sum_{j=0}^{n}P_{n,j} \vec{z}_j\ . 
\label{eq:z2x}
\end{equation}
On the left side appears the expression which is obtained as the result of the measurement, the right hand side contains the parameters to be fitted, the magnet offsets and angles and the initial orbit coordinates. 
Since there are more parameters than measured values, this expression can only be solved by fitting at least two different measurements with different quadrupole settings (index $m$, thus different matrices $T_{n,0}^{(m)}$ and $P_{n,j}^{(m)}$) simultaneously, or adding as additional constraint that the magnet positions should differ as little as possible from their nominal value $\vec{z}_n=\vec{z}_{n}^0$.  Additional constraints can be the readings of the beam position monitors BPMs in the IR region.
\par
From the previous paragraph, it is clear that the angular alignment cannot be determined with satisfactory precision. Since the angles of the magnets with respect to the beam need to be taken into account for the fit, the design angles of the magnet are used.  Therefore we can only make use of the position part of the vector equation (\ref{eq:z2x}).  This introduces certain errors into the alignment reconstruction that are analyzed in the next section.

We now define a new relationship between the measured values of $x_n-z_n$, the magnet offsets $z_0, z_1\ldots z_N$ and the initial orbit values $x_0,x'_0$.  For this we define new vectors.  The first one includes several sets
 $(m\in \{1\ldots M\})$ of relative position measurements $(x_m-z_m)^{(m)}$ in all the IR magnets, as well as the design position $z_n^0$
\begin{equation}
\vec{v}=((x_0-z_0)^{(1)}, (x_1-z_1)^{(1)}, \ldots (x_N-z_N)^{(1)},
\ldots (x_N-z_N)^{(M)}, z_0^0,\ldots z_N^0)\ .
\end{equation}
The second vector contains the parameters to be determined
\begin{equation}
\vec{g}=(z_0,\ldots z_N,x_0^{(1)}, x_0^{'(1)},\ldots x_0^{(M)}, x_0^{'(M)})\ ,
\end{equation}
and the third vector $\vec{w}$ contains the parameters which are kept fixed, the  design angles of the magnets $z_j^{'0}$ and the effects of the dipole corrector settings of the $m$th measurement,
\begin{eqnarray}
w_n^{(m)} &=&
\left[T_{n,0}^{(m)}\right]_{1,5}
+\sum_{j=0}^n \left[P_{0,j}^{(m)}\right]_{1,2}\cdot z_j^{'0}
\ {\rm for}\ j\in\{1,\ldots N\cdot M\}\ ,\nonumber\\
w_n^{(m)} &=& 0 \ {\rm else}
\ .
\end{eqnarray}
Note the indices outside the square brackets denote the matrix element,
the indices inside the square bracket denote the matrix.
The measurements are then related to the parameters by
\begin{equation}
\vec{v}= A\vec{g}+\vec{w}\ .
\end{equation}
The matrix A contains the matrix elements which are determined by equation (\ref{eq:z2x}),
\begin{eqnarray}
A_{(m-1)\cdot N+n,j}&=&\left[P_{n,j}^{(m)}\right]_{1,1} \ {\rm for}\ 0\le j\le n\ ,\\
A_{(m-1)\cdot N+n,j}&=&\left[T_{n,0}^{(m)}\right]_{1,1} \ {\rm for}\ j=N+2\cdot (m-1)+1\ ,\\
A_{(m-1)\cdot N+n,j}&=&\left[T_{n,0}^{(m)}\right]_{1,2} \ {\rm for}\ j=N+2\cdot (m-1)+2\ ,\\
A_{M\cdot N+j,j} &=& 1\ {\rm for}\ 0<j\le N\ ,\\
A_{(m-1)\cdot N+n,j}&=&0\ {\rm else}\ .
\end{eqnarray}
The solution of the fit with a quadratic norm  is 
\begin{equation}
\vec{g}= (A^T A)^{-1} A^T  (\vec{v}-\vec{w})\ .
\label{eqn49}
\end{equation}
In cases when the inverse cannot be computed due to a bad condition of the matrix, a singular value decomposition (SVD) can be helpful.  

\subsection{Error in Magnet Offset Determination by Ignoring the Magnet Angle Offsets}
The quadrupole magnet's angular alignment produces a contribution to the closed orbit that is to be reconstructed. Since we are not able to measure the angular alignment, it is not contained in our model. 
Therefore, the missing angular offset in the model is compensated by an additional false offset in neighboring magnets.  The magnitude of this error is estimated in the following.
\par
The effect of the magnet angle on the beam trajectory is described in equation (\ref{eq:xproper}) for a one of the orbit planes by two dimensional sub-matrices,
\begin{eqnarray}
x_n&=& T_{nj}Q_j
\left(\begin{array}{c} 0\\ z'_j-z_j^{' 0}\end{array}\right)\ ,\\
Q_j
&=&
g_j^{-1} t_j-g_j t_j^{-1}\\
&=&
\left(\begin{array}{cc}
0 & 2\frac{\frac{l_j}{2}\sqrt{k_j}\cos(\frac{l_j}{2}\sqrt{k_j})-\sin(\frac{l_j}{2}\sqrt{k_j})}{\sqrt{k_j}}\\
2\sqrt{k_j}\sin(\frac{l_j}{2}\sqrt{k_j}) & 0
\end{array}\right)\ ,\nonumber\\
x_n&=& T_{nj}\left(\begin{array}{c}1\\ 0\end{array}\right)
2\frac{\frac{l_j}{2}\sqrt{k_j}\cos(\frac{l_j}{2}\sqrt{k_j})-\sin(\frac{l_j}{2}\sqrt{k_j})}{\sqrt{k_j}}
(z'_j-z_j^{' 0})\ .\label{eq:misang}
\end{eqnarray}
We use the same notation as before, $T_{nj}$ is the transport matrix between the middle of quadruple index $j$ and $n$,  $g_j$ is the transformation through half of this quadrupole and $t_j$ is a transformation though the drift of the same half length, $z'_j$ is the angle of the beam with respect to the design curve. In our model, the missing effect of the angle alignment in equation (\ref{eq:misang}) is produced by additional, false magnet offsets of neighboring quadrupoles.  Quadrupole $i$ requires an additional shift $\Delta z_i$ and an additional angle $\Delta z_i'$ to produce the effect of the angle of quadrupole $j$,
\begin{equation}
x_n = T_{ni}Q_i
\left(\begin{array}{l} \Delta z_i\\ \Delta z_i'\end{array}\right)\ .
\end{equation}
These false offsets are given by 
\begin{equation}
\left(\begin{array}{l} \Delta z_i\\ \Delta z_i'\end{array}\right)
=
Q_i^{-1}T_{ni}^{-1}T_{nj}
\left(\begin{array}{c}1\\0\end{array}\right)
2\frac{\frac{l_j}{2}\sqrt{k_j}\cos(\frac{l_j}{2}\sqrt{k_j})-\sin(\frac{l_j}{2}\sqrt{k_j})}{\sqrt{k_j}}
(z'_j-z_j^{' 0})\ .
\end{equation}
For two quadrupoles with a phase distance of $\phi_{ij}=-\arctan(1/\alpha_j)$, taking the atan function in $\{0,\pi]$, one obtains
\begin{equation}
T_{ni}^{-1}T_{nj}=T_{ij}=
\left(\begin{array}{cc}
0&-\sqrt{\frac{\beta_i}{\gamma_j}}\\
\sqrt{\frac{\gamma_j}{\beta_i}}&
\frac{\alpha_i-\alpha_j}{\sqrt{\gamma_j\beta_i}}
\end{array}\right)\ .
\end{equation}
This leads to $\Delta z_i'=0$ so that the quadrupole $i$ alone can compensate the missing angle of quadrupole $j$.  The error of the position reconstruction for this quadrupole is given by
\begin{equation}
\Delta z_i=
\sqrt{\frac{\gamma_j}{\beta_i}}
\frac{\frac{l_j}{2}\sqrt{k_j}\cos(\frac{l_j}{2}\sqrt{k_j})-\sin(\frac{l_j}{2}\sqrt{k_j})}{\sqrt{k_i k_j}\sin(\frac{l_i}{2}\sqrt{k_i})}
(z'_j-z_j^{' 0})\ .
\end{equation}
The factor between $\Delta z_i$ and $z'_j-z_j^{' 0}$ depends very much on the values of $\alpha$ in the centers of the two quadrupoles an on the chosen phase advance which allows that a single quadrupole can compensate the ignored angular alignment of magnet $j$.

In a realistic setting, at least two quadrupole positions, $i$ and $q$, will be reconstructed erroneously to take account of the missing angle alignment of quadrupole $j$.  For this the following equation has to be satisfied,
\begin{equation}
T_{ni}Q_i\left(\begin{array}{c}\Delta z_i\\ 0\end{array}\right)
+
T_{nq}Q_q\left(\begin{array}{c}\Delta z_q\\ 0\end{array}\right)
=
T_{nj}Q_j\left(\begin{array}{c}0\\ z_j'-z_j^{' 0}\end{array}\right)
\end{equation}
Solving for the position errors leads to
\begin{equation}
\left(\begin{array}{l}\Delta z_i\\ \Delta z_q\end{array}\right)
=
(z_j'-z_j^{' 0})
\frac{ \frac{l_j}{2}\sqrt{k_j} \cos(\frac{l_j}{2}\sqrt{k_j})-\sin(\frac{l_j}{2}\sqrt{k_j})}{\sin\phi_{qi}\sqrt{\beta_j k_j}}
\left(\begin{array}{r}
\frac{\cos\phi_{jq}-\alpha_j\sin\phi_{jq}}{\sin(\frac{l_i}{2}\sqrt{k_i})\sqrt{\beta_i k_i}}
\\
-\frac{\cos\phi_{ji}-\alpha_j\sin\phi_{ji}}{\sin(\frac{l_q}{2}\sqrt{k_q})\sqrt{\beta_q k_q}}
\end{array}\right)\ .
\end{equation}

Table \ref{tb:alignproper} shows what error an angular alignment error of HERA's IR quadrupoles can have on the estimated position of the two neighboring IR magnets.  Even for an angular alignment error of 1mrad, the errors in the reconstructed position is always below $150\mu$m.
\begin{table}
\begin{center}
\caption{Effect of an angular alignment error of HERA IR magnets on the estimated alignment of the two neighboring quadrupoles.  Since the nominal quadrupole strength of GGR is zero, this magnet is not considered here.\label{tb:alignproper}}
\begin{tabular}{lcc}
\hline
Name & Quad to the left & Quad to the right \\
     & mm/mrad          & mm/mrad           \\
\hline
QL14L &\phantom{-}0.000 &          -0.001 \\
GJ8L  &\phantom{-}0.029 &          -0.017 \\
GI7L  &\phantom{-}0.074 &          -0.056 \\
GOL   &\phantom{-}0.104 &\phantom{-}0.050 \\
GI6R  &          -0.016 &          -0.041 \\
GI7R  &\phantom{-}0.070 &          -0.138 \\
GJ8R  &\phantom{-}0.014 &          -0.030 \\
QL14R &\phantom{-}0.001 &\phantom{-}0.000 \\
\hline
\end{tabular}
\end{center}
\end{table}
\subsection{Error Propagation in the Fitting Procedure}
The magnet positions as the result of a fit of the measurements
can be written as 
\begin{equation}
z_n=\sum_{j=1}^N B_{nj}y_j\ .
\end{equation}
The $z_n$ are the first $N$ components of the vector $\vec{g}$, the  $y_j$ are the components of $\vec{v}-\vec{w}$ and $B=(A^TA)^{-1}A^T$ is the matrix of the least square fit in equation (\ref{eqn49}).
We now introduce sets of random errors of the input variables labeled by $\alpha$ 
\begin{equation}
z_n+\Delta z_n^\alpha
=\sum_{j=1}^N (B_{nj}+\Delta B_{n,j}^\alpha)(y_j+\Delta y_j^\alpha)\ .
\end{equation}
which gives the error of the magnet positions (neglecting second order terms)
\begin{equation}
\Delta z_n^\alpha=\sum_{j=1}^N (\Delta B_{n,j}^\alpha\cdot y_j+
B_{n,j}\cdot \Delta y_j^\alpha)\ .
\end{equation}
We now calculate the expectation value of $\Delta z_n$ by squaring the expression and by subsequently averaging over the error set $\alpha$.
We assume that for random errors the following correlations hold:
$<\Delta B_{n,j}^\alpha\Delta B_{nk}^\alpha>_\alpha
=\Delta B_{{\rm rms}}^2\delta_{j,k}$ and 
$<\Delta y_j^\alpha\Delta y_k^\alpha>_\alpha=y_{{\rm rms}}^2\delta_{j,k}$.
With this we finally obtain
\begin{equation}
\Delta z_{n,rms}=\sqrt{
\sum_j y_{j}^2\cdot\Delta B_{{\rm rms}}^2+\sum_{j} B_{n,j}^2\cdot\Delta y_{{\rm rms}}^2
}\ .
\end{equation}
The first sum has typically values of $1$ to $3$ when evaluated for the 
HERA IR. Thus an error of a single measurement $\Delta(x_j-z_j)=0.1$mm propagates, yielding approximately an error of $0.3$mm in the reconstructed magnet position.  However, these are only first observations for the case of HERA where the here presented version of beam-based alignment is currently being heavily used in the commissioning process.  This particular application and experiences with this method will thus be reported in a separate paper after the successful commissioning of the HERA luminosity upgrade.

\section{Conclusion}
We have introduced a beam based alignment method for a general class of combined function magnets that can be encountered in collider interaction regions.  While this method can in principal determine alignment angles, we have shown that these angles would be very prone to measurement errors.  We have therefore introduced a procedure to use the angular alignment to strongly improve the accuracy of the position determination by one to two orders of magnitude.  Furthermore a procedure has been presented to determine global magnet positions after the closed orbit deviation from the quadrupole axis has been measured throughout the interaction region.


\end{document}